\documentclass[12pt]{article}
\usepackage{mathrsfs} \usepackage{amsmath} \usepackage{amssymb} \usepackage{slashed}
\usepackage{graphicx} \usepackage{caption2}
\begin{document}
\newcommand{\ba}{\begin{eqnarray}} \newcommand{\ea}{\end{eqnarray}}
\newcommand{\be}{\begin{equation}} \newcommand{\ee}{\end{equation}}
\renewcommand{\figurename}{Figure.} \renewcommand{\captionlabeldelim}{.~}
\renewcommand{\thefootnote}{\fnsymbol{footnote}}

\vspace*{1cm}
\begin{center}
 {\Large\textbf{Dark Matter, Leptogenesis, and Neutrino Mass in the Minimal Extension of the SM with $U(1)_{B-L}\otimes Z_{2}$}}

 \vspace{1cm}
 \textbf{Wei-Min Yang}

 \vspace{0.3cm}
 \emph{Department of Modern Physics, University of Science and Technology of China, Hefei 230026, P. R. China}

 \emph{E-mail: wmyang@ustc.edu.cn}
\end{center}

\vspace{1cm}
\noindent\textbf{Abstract}: I suggest a minimal extension of the SM with $U(1)_{B-L}\otimes Z_{2}$. It can simultaneously accommodate the tiny neutrino mass, cold dark mater and baryon asymmetry besides the SM. All of the new physics arise from the $U(1)_{B-L}$ violation at the energy scale about 1000 TeV. The model can naturally explain the ``WIMP Miracle" and elegantly achieve the leptogenesis, in addition, some predictions of the model are possible and promising to be tested in near future experiments.

\vspace{1cm}

\newpage
\noindent\textbf{I. Introduction}

\vspace{0.3cm}
  The standard model (SM) of the particle physics has successfully accounted for all kinds of the physics at or below the electroweak scale \cite{1}, but it can not explain some important issues such as the tiny neutrino mass \cite{2}, the cold dark matter (CDM) \cite{3} and the baryon asymmetry \cite{4}. Some inspired theoretical ideas have been suggested to solve these problems, for instance, the tiny neutrino mass can be generated by the seesaw mechanism \cite{5a} or the other means \cite{5b}, the baryon asymmetry can be achieved by the thermal leptogenesis \cite{6} or the electroweak baryogenesis \cite{7}, the CDM candidates are possibly the sterile neutrino \cite{8}, the lightest supersymmetric particle \cite{9}, the axion \cite{10}, and so on. Some possible connections among the neutrino mass, the CDM, and the baryon asymmetry have been discussed in many references, for example, the neutrino mass and the leptogenesis are implemented by the heavy Higgs triplets \cite{11a}, the asymmetric CDM is related to the baryon asymmetry \cite{11b}, and some new models \cite{11c}. Although many progresses on these fields have been made all the time \cite{11d}, a convincing and unified theory is not established as yet.

  If we believe the universe harmony and the nature unification, it is hard to believe that the tiny neutrino mass, the CDM and the matter-antimatter asymmetry appear to be not related to each other. It is very possible that the three things have a common origin, therefore, a realistic theory beyond the SM should simultaneously account for the three things, moreover, it should keep such principles as the simplicity and the fewer number of parameters. In addition, the new theory should be feasible and promising to be tested in future experiments. If one theory is excessive complexity and unable to be tested, it is unbelievable and infeasible. Based on these considerations, I suggest a minimal extension of the SM. It only adds the symmetries of $U(1)_{B-L}\otimes Z_{2}$ and a few particles to the SM but it is able to accommodate the above three things, in addition, we are likely to probe the attractive new physics at the future colliders.

\vspace{0.6cm}
\noindent\textbf{II. Model}

\vspace{0.3cm}
  The model only introduces two symmetries of the local gauge $U(1)_{B-L}$ and a discrete $Z_{2}$ besides the SM gauge groups $G_{SM}$, where $B$ and $L$ respectively denote the baryon and lepton numbers. The model particle contents and their gauge quantum numbers under $SU(2)_{L}\otimes U(1)_{Y}\otimes U(1)_{B-L}$ are listed as follows,
\begin{alignat}{1}
 &l_{L}(2,-1,-1)_{1},\hspace{0.5cm} e^{-}_{R}(1,-2,-1)_{1},\hspace{0.5cm} N^{0}_{R}(1,0,-1)_{-1}, \nonumber\\
 &H(2,1,0)_{1},\hspace{0.5cm} \Delta(3,-2,-2)_{1},\hspace{0.5cm} \phi^{0}(1,0,-2)_{1},\hspace{0.5cm} \phi^{-}(1,-2,-2)_{-1},
\end{alignat}
where the right subscripts of the brackets are the particle parities under the $Z_{2}$ transformation, which means $N_{R}\rightarrow-N_{R}$, $\phi^{-}\rightarrow-\phi^{-}$, and the rest of the fields are transformed into themselves. The $Z_{2}$ symmetry will lead that $N_{R}$ become the Majorana fermion and the CDM. In addition, $N_{R}$ has three generations as the other fermions so that the chiral anomaly is completely cancelled, therefore the model is anomaly-free. In (1) I ignore the quark sector and the color subgroup $SU(3)_{C}$ because what followed have nothing to do with them.

  The invariant Lagrangian of the model under the above symmetries is composed of the three following parts. The gauge kinetic energy terms are
\begin{alignat}{1}
 \mathscr{L}_{G}&=\mathscr{L}_{pure\:gauge}+\sum\limits_{f}i\,\overline{f}\,\gamma^{\mu}D_{\mu}f+\sum\limits_{S}(D^{\mu}S)^{\dagger}D_{\mu}S\,, \nonumber\\
 D_{\mu}&=\partial_{\mu}+i\left(g_{2}W_{\mu}^{i}\frac{\tau^{i}}{2}+g_{1}B_{\mu}\frac{Y}{2}+g_{0}X_{\mu}\frac{B-L}{2}\right),
\end{alignat}
where $f$ and $S$ respectively denote all kinds of the fermions and scalars in (1), $X_{\mu}$ is the gauge field associated with $U(1)_{B-L}$, the other notations are self-explanatory. The Yukawa couplings are
\ba
 \mathscr{L}_{Y}=\overline{l_{L}}HY_{E}e_{R}+\frac{1}{2}l_{L}^{T}C\Delta^{\dagger}Y_{\nu}l_{L}+\frac{1}{2}\phi^{0*}N_{R}^{T}CY_{N}N_{R}+\phi^{+}N_{R}^{T}CY_{D}e_{R}+h.c.\,,
\ea
where $C$ is a charge conjugation matrix. Note that the $Z_{2}$ symmetry forbids the two couplings of $\overline{l_{L}}i\tau_{2}H^{*}YN_{R}$ and $\phi^{+}l_{L}^{T}CYi\tau_{2}l_{L}$ even though they satisfy the gauge symmetries, in which $\tau_{2}$ is the second Paul matrix and it is inserted to satisfy the $SU(2)_{L}$ symmetry. This will ensure the CDM stability. The coupling parameters $Y_{E,\nu,N,D}$ are generally $3\times3$ complex matrices. However we can choose such flavour basis in which $Y_{E}$ and $Y_{N}$ are simultaneously diagonal matrices, thus some phases in $Y_{\nu}$ and $Y_{D}$ which can not be removed by the redefined field phases become $CP$-violating sources in the lepton sector in comparison with one in the quark sector. After $U(1)_{B-L}$ breaking, the second and third terms in (3) will give rise to the light neutrino mass and the CDM one, respectively, while the last term plays a key role in the CDM annihilation and the leptogenesis.

  The full scalar potentials are
\begin{alignat}{1}
 V_{S}=&\frac{1}{4\lambda_{\Delta}}\left(2\lambda_{\Delta}Tr[\Delta^{\dagger}\Delta]-((\lambda_{\Delta}+\lambda_{0})v_{\Delta}^{2}+\lambda_{1}v_{H}^{2}+\lambda_{2}v_{\phi}^{2})+(\frac{M_{\Delta}v_{\Delta}}{v_{\Delta}})^{2}\right)^{2} \nonumber\\
 &+\frac{1}{4\lambda_{H}}\left(2\lambda_{H}H^{\dagger}H-(\lambda_{1}v_{\Delta}^{2}+\lambda_{H}v_{H}^{2}+\lambda_{3}v_{\phi}^{2})+(\frac{M_{\Delta}v_{\Delta}}{v_{H}})^{2}\right)^{2} \nonumber\\
 &+\frac{1}{4\lambda_{\phi^{0}}}\left(2\lambda_{\phi^{0}}\phi^{0*}\phi^{0}-(\lambda_{2}v_{\Delta}^{2}+\lambda_{3}v_{H}^{2}+\lambda_{\phi}v_{\phi}^{2})+(\frac{M_{\Delta}v_{\Delta}}{v_{\phi}})^{2}\right)^{2} \nonumber\\
 &+\frac{1}{4\lambda_{\phi^{-}}}\left(2\lambda_{\phi^{-}}\phi^{+}\phi^{-}-(\lambda_{4}v_{\Delta}^{2}+\lambda_{5}v_{H}^{2}+\lambda_{6}v_{\phi}^{2})+M^{2}_{\phi^{-}}\right)^{2} \nonumber\\
 &+\lambda_{0}Tr[(\Delta^{\dagger}\Delta)^{2}]+2\lambda_{1}Tr[\Delta^{\dagger}\Delta]H^{\dagger}H+2\left(\lambda_{2}Tr[\Delta^{\dagger}\Delta]+\lambda_{3}H^{\dagger}H\right)\phi^{0*}\phi^{0} \nonumber\\
 &+2\left(\lambda_{4}Tr[\Delta^{\dagger}\Delta]+\lambda_{5}H^{\dagger}H+\lambda_{6}\phi^{0*}\phi^{0}\right)\phi^{+}\phi^{-} \nonumber\\
 &+2\left(\lambda_{7}Tr[H^{T}i\tau_{2}\Delta i\tau_{2}H]\phi^{0*}+\lambda_{8}Tr[\Delta i\tau_{2}\Delta i\tau_{2}]\phi^{+}\phi^{+}+h.c.\right),
\end{alignat}
where $v_{\Delta}=\frac{\lambda_{7}v_{H}^{2}v_{\phi}}{M_{\Delta}^{2}}$ is not an independent parameter, in fact, there are only four independent mass-dimensional parameters, namely $[M_{\Delta},v_{H},v_{\phi},M_{\phi^{-}}]>0$, in which $M_{\Delta}$ and $M_{\phi^{-}}$ are respectively the original masses of $\Delta$ and $\phi^{-}$. All of the self-coupling parameters satisfy the requirements as $[\lambda_{\Delta},\lambda_{H},\lambda_{\phi^{0}},\lambda_{\phi^{-}},\lambda_{0}]\sim0.1>0$, while all of the interactive coupling parameters are assumed as $[\lambda_{1},\lambda_{2},\ldots,\lambda_{8}]\ll 1$, in other words, the interactions among the scalars are very weak. These conditions are natural and reasonable, however, they can sufficiently guarantee the vacuum stability. The vacuum configurations derived from the $V_{Scalar}$ minimum are exactly
\ba
 \langle\Delta\rangle =\left(\begin{array}{cc}\frac{v_{\Delta}}{\sqrt{2}}&0\\0&0\end{array}\right),\hspace{0.5cm} \langle H\rangle =\left(\begin{array}{c}0\\\frac{v_{H}}{\sqrt{2}}\end{array}\right),\hspace{0.5cm} \langle\phi^{0}\rangle=\frac{v_{\phi}}{\sqrt{2}}\,,\hspace{0.5cm} \langle\phi^{-}\rangle=0\,.
\ea
The vacuum expectation values and the mass parameters are assumed to be such hierarchy as
\ba
v_{\Delta}\ll v_{H}\approx246\:\mathrm{GeV}\sim M_{\phi^{-}}\ll v_{\phi}\sim2000\:\mathrm{TeV}\sim M_{\Delta}.
\ea
$v_{H}$ has been fixed by the electroweak physics. $v_{\Delta}$ can be determined by the tiny neutrino mass. $v_{\phi}$ and $M_{\phi^{-}}$ will be determined by the leptogenesis and the CDM, respectively. It is more natural and reasonable in the model that $M_{\Delta}$ is close to $v_{\phi}$ rather than a super-high scale, the reason for this is that a large hierarchy between it and $v_{\phi}$ is unnatural, moreover, we don't need the super-high scale physics.

  $\langle\phi^{0}\rangle$ early violates the local $U(1)_{B-L}$, later $\langle H\rangle$  results in the spontaneous breaking of $SU(2)_{L}\otimes U(1)_{Y}$. $\langle\Delta\rangle$ simultaneously breaks the both, but $v_{\Delta}$ is so small that it has effectively no impacts on the two breakings. However, the $Z_{2}$ symmetry is always kept. After gauge symmetry breakings the particles obtain their masses as follows,
\begin{alignat}{1}
 &M_{W_{\mu}}=\frac{v_{\phi}g_{2}}{2}\,,\hspace{0.3cm} M_{Z_{\mu}}=\frac{v_{\phi}g_{2}}{2cos\theta_{w}}\,,\hspace{0.3cm} M_{X_{\mu}}=v_{\phi}g_{0},\hspace{0.3cm} M_{H^{0}}= v_{H}\sqrt{2\lambda_{H}}\,,\hspace{0.3cm} M_{\phi^{0}}=v_{\phi}\sqrt{2\lambda_{\phi^{0}}}\,, \nonumber\\
 &M_{e}=-\frac{v_{H}}{\sqrt{2}}Y_{E},\hspace{0.5cm} M_{\nu}=-\frac{v_{\Delta}}{\sqrt{2}}Y_{\nu}=-\frac{v_{H}^{2}v_{\phi}}{\sqrt{2}M_{\Delta}^{2}}\lambda_{7}Y_{\nu},\hspace{0.5cm} M_{N}=-\frac{v_{\phi}}{\sqrt{2}}Y_{N},
\end{alignat}
where $tan\theta_{w}=\frac{g_{1}}{g_{2}}$ is the weak mixing angle. Note that the mixing between $Z_{\mu}$ and $X_{\mu}$ is nearly zero due to $\frac{v_{\Delta}^{2}}{v_{\phi}^{2}}\approx0$, in addition, the mixing between $H^{0}$ and $\phi^{0}$ is very weak since $\lambda_{3}$ is very small. The masses of the SM particles have all been measured. $M_{X_{\mu}}$, $M_{\phi^{0}}$, and $M_{N}$ beyond the SM are all proportional to $v_{\phi}$, so they are probably $\sim1000$ TeV or so, but the lightest one of $M_{N}$ is possibly several hundred GeVs on account of the large hierarchy of $Y_{N}$. In view of $\frac{v_{H}^{2}v_{\phi}}{M_{\Delta}^{2}}\sim10^{-2}$ GeV, the tiny neutrino mass is correctly generated provided $\lambda_{7}Y_{\nu}\sim 10^{-9}$, furthermore, the experimental data of the neutrino masses and mixing angles can accurately be fitted by choosing suitable texture of $Y_{\nu}$.

  Finally, we can assume such spectrum relation for the three mass eigenvalues of $M_{N}$ and the two scalar masses $M_{\phi^{0}}, M_{\phi^{-}}$ as follows,
\ba
 M_{Z}<M_{N_{1}}<M_{\phi^{-}}<M_{Z}+M_{N_{1}}\sim(200-500)\:\mathrm{GeV}\ll M_{\phi^{0}}\lesssim M_{N_{2}}\lesssim M_{N_{3}}\sim v_{\phi}\,.
\ea
This is easily  satisfied by some suitable values of the coupling parameters in (7), so (8) is however reasonable and believable. These mass restrictions will lead to the successful CDM and leptogenesis.

\vspace{0.6cm}
\noindent\textbf{III. Cold Dark Matter}

\vspace{0.3cm}
  In the light of the model couplings, the heavier $N_{2,3}$ can decay as $N_{2,3}\rightarrow\overline{e}_{\alpha}+\phi^{-}$, but the lightest $N_{1}$ can not decay due to the kinetic restriction of (8). Therefore $N_{1}$ is a stable particle in the model, moreover, it is really a WIMP. The natures of $N_{1}$ are very well consistent with ones of the CDM, so it becomes a desirable candidate of the CDM.

  The current relic abundance of $N_{1}$ can be calculated by the thermal production in the early universe. After $N_{1}$ becomes non-relativistic particle, it has two annihilation channels, (i) the pair annihilation $N_{1}+\overline{N}_{1}\rightarrow\overline{e}_{\alpha}+e_{\beta}$ via the t-channel mediation of $\phi^{-}$, shown as (a) in Figure 1, (ii) the associated annihilation $Z_{\mu}^{0}+N_{1}\rightarrow\overline{e}_{\alpha}+\phi^{-}$, shown as (b) in Figure 1.
\begin{figure}
 \centering
 \includegraphics[totalheight=5cm]{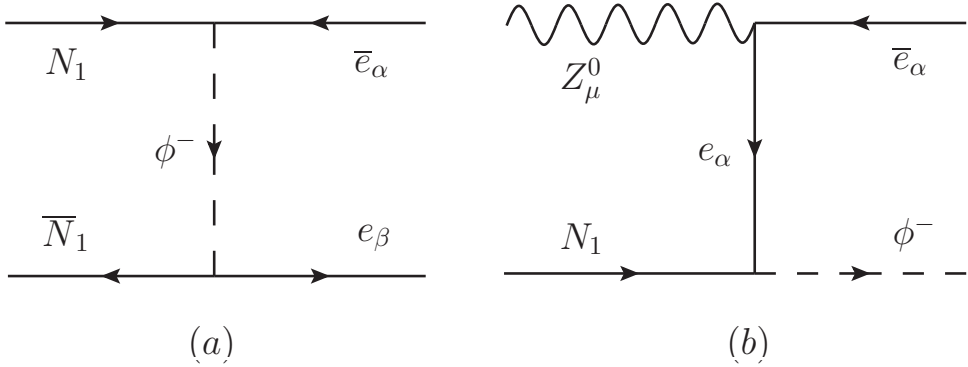}
 \caption{(a) The pair annihilation of the CDM $N_{1}$. (b) The associated annihilation of the CDM $N_{1}$ with $Z^{0}_{\mu}$, which determines the free-out temperature and relic abundance of the CDM $N_{1}$.}
\end{figure}
Note that the process $\gamma+N_{1}\rightarrow\overline{e}_{\alpha}+\phi^{-}$ won't work due to (8), and the process $W_{\mu}^{-}+N_{1}\rightarrow\overline{\nu}_{\alpha}+\phi^{-}$ can't take place because $e_{\alpha}$ in the propagator must be a right-handed lepton. After some careful analysis, the cross-section of (i) is much smaller than one of (ii) because (i) has the heavy propagator with $M_{\phi^{-}}$, therefore $N_{1}$ decoupling and its relic abundance are essentially dominated by the (ii) process. The annihilation rate of (ii) is calculated as follows,
\begin{alignat}{1}
 &\Gamma=\sum\limits_{\alpha}\Gamma(N_{1}+Z_{\mu}^{0}\rightarrow \overline{e}_{\alpha}+\phi^{-})=\langle\sigma v \rangle n_{N_{1}}, \nonumber\\
 &\langle\sigma v \rangle=a+b\,\langle v^{2}\rangle+c\,\langle v^{4}\rangle+\cdots,\hspace{0.5cm} n_{N_{1}}=2\left(\frac{M_{N_{1}}T}{2\pi}\right)^{\frac{3}{2}}e^{-\frac{M_{N_{1}}}{T}}, \nonumber\\
 &\sigma v=\frac{\alpha_{e}tan^{2}\theta_{w}(Y_{D}Y_{D}^{\dagger})_{11}}{6M^{2}_{Z}}(1-\frac{M^{2}_{\phi^{-}}}{s})\left(1-(\frac{M^{2}_{N_{1}}-M^{2}_{Z}}{s})^{2}\right)^{-1}(A_{1}+A_{2}+A_{3}), \nonumber\\
 &A_{1}=\frac{M^{4}_{Z}}{t_{0}t_{1}}(\frac{M^{2}_{\phi^{-}}-M^{2}_{N_{1}}}{s})\,,\hspace{0.3cm} A_{2}=\frac{2M^{2}_{Z}}{t_{0}+t_{1}}(\frac{M^{2}_{N_{1}}}{s}-1)\,,\hspace{0.3cm} A_{3}=\frac{1}{4}(1-\frac{M^{2}_{\phi^{-}}}{s})(1+\frac{M^{2}_{N_{1}}-M^{2}_{Z}}{s}),\nonumber\\
 &t_{0}(t_{1})=(E_{N_{1}}-E_{\phi^{-}})^{2}-(p_{N_{1}}\mp p_{\phi^{-}})^{2},\nonumber\\
 &v=\frac{2s\sqrt{(s-(M_{N_{1}}+M_{Z})^{2})(s-(M_{N_{1}}-M_{Z})^{2})}}{s^{2}-(M^{2}_{N_{1}}-M^{2}_{Z})^{2}}\,,\hspace{0.5cm} \langle v^{2}\rangle=\frac{3T}{M_{N_{1}}}+\frac{3T}{M_{Z}}\,,
\end{alignat}
where $\alpha_{e}\approx\frac{1}{129}$ and $tan\theta_{w}\approx0.55$. $v$ is a relative velocity of two annihilating particles. $s$ and $t$ are the Mandelstam variables. $E_{N_{1}},E_{\phi^{-}}$ and $p_{N_{1}},p_{\phi^{-}}$ are the particle energies and momenta in the center-of-mass frame, which are all functions of $s$. Note that $v=0$ corresponds to $s=(M_{N_{1}}+M_{Z})^{2}$. $n_{Z}$ is larger than $n_{N_{1}}$ due to $M_{Z}<M_{N_{1}}$, so the pairing number density of the annihilating $Z_{\mu}$ and $N_{1}$ is namely equal to $n_{N_{1}}$. In view of (8), the momentum transfer $t_{0}(t_{1})$ is smaller than $M^{2}_{Z}$ since $Z_{\mu}$ and $N_{1}$ are non-relativistic states at the decoupling temperature. This leads that the leading term $A_{1}$ of $\sigma v$ is roughly $\sim10^{2}$ in (9), then $(Y_{D}Y_{D}^{\dagger})_{11}A_{1}\sim 1$ is very reasonable, thus we can naturally obtain $\sigma v\sim10^{-9}$ $\mathrm{GeV}^{-2}$, which is exactly a weak interaction cross-section. The thermal average of the annihilation cross-section at the freeze-out temperature is close to a weak interaction cross-section, this is so-called ``WIMP Miracle" \cite{3}.

  $N_{1}$ is decoupling when its annihilation rate is equal to the Hubble expansion rate of the universe, the freeze-out temperature is thus determined by
\begin{alignat}{1}
 &\Gamma(T_{f})=H(T_{f})=\frac{1.66\sqrt{g_{*}(T_{f})}\,T_{f}^{2}}{M_{Pl}}\,, \nonumber\\
\Longrightarrow &x=\frac{T_{f}}{M_{N_{1}}}\approx\left(17.6+ln\frac{M_{N_{1}}}{\sqrt{g_{*}(T_{f})x}}+ln\frac{\langle\sigma v \rangle}{10^{-10}\:\mathrm{GeV^{-2}}}\right)^{-1},
\end{alignat}
where $M_{Pl}=1.22\times10^{19}$ GeV, $g_{*}$ is the effective number of relativistic degrees of freedom. The current relic abundance of $N_{1}$ is calculated by the following equation \cite{12},
\ba
 \Omega_{N_{1}}h^{2}=\frac{0.85\times10^{-10}\:\mathrm{GeV}^{-2}}{\sqrt{g_{*}(T_{f})}\,x(a+3b\,x)}\approx 0.12,
\ea
where $a$ and $b$ are determined by (9). $0.12$ is the current abundance of the CDM \cite{13}. Obviously, $M_{N_{1}}$ and $\langle\sigma v\rangle$ are in charge of the final results of (10) and (11). For $M_{N_{1}}\sim300$ GeV and $\langle\sigma v\rangle\sim10^{-9}$ $\mathrm{GeV}^{-2}$, the solution of (10) is $x\sim\frac{1}{27}$, then $T_{f}\sim 11$ GeV. At this temperature the relativistic particles include $photon,gluon,\nu,e^{-},\mu^{-},u,d,s$, so we can figure out $g_{*}(T_{f})=61.75$ in (10) and (11). Finally, we can correctly reproduce $\Omega_{N_{1}}h^{2}\approx0.12$ by use of (11). In conclusion, the model can simply and naturally account for the CDM, in particular, explain the ``WIMP Miracle".

\vspace{0.6cm}
\noindent\textbf{IV. Leptogenesis}
\vspace{0.3cm}

  The model can also account for the baryon asymmetry through the leptogenesis at the scale about a thousand TeVs. After the $B-L$ symmetry is broken, the massive $\phi^{0}$ has three decay modes on the basis of the model, (i) the two-body decay $\phi^{0}\rightarrow N_{1}+N_{1}$, (ii) the three-body decay $\phi^{0}\rightarrow N_{1}+\overline{e}_{\alpha}+\phi^{-}$, Figure 2 shows its tree and loop diagrams,
\begin{figure}
 \centering
 \includegraphics[totalheight=8cm]{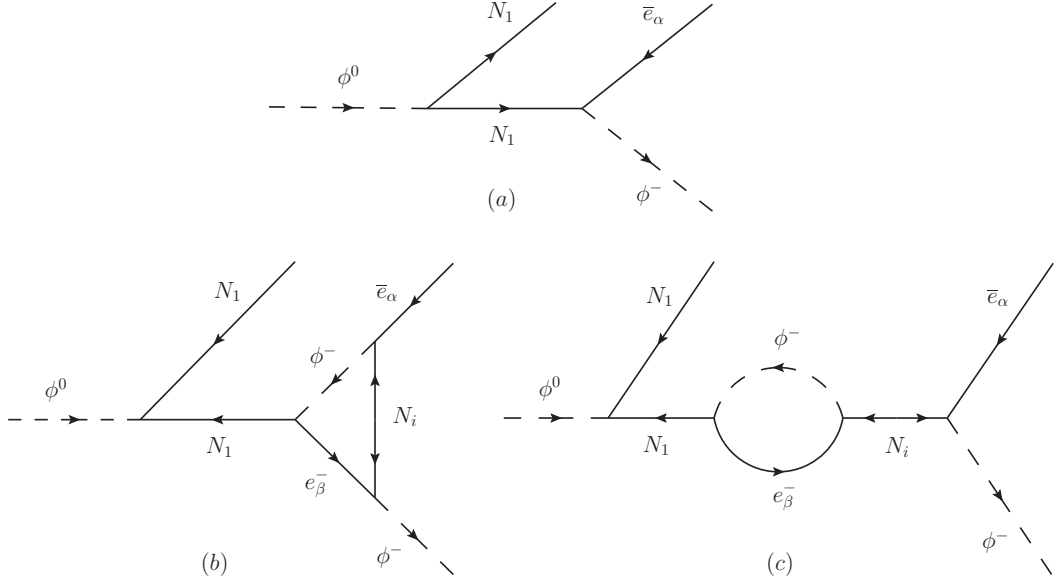}
 \caption{The tree and loop diagrams of the $B-L$ violating decay $\phi^{0}\rightarrow N_{1}+\overline{e}_{\alpha}+\phi^{-}$, which is out-of-equilibrium and $CP$-asymmetric decay, it eventually leads to the matter-antimatter asymmetry.}
\end{figure}
(iii) the radiative decay $\phi^{0}\rightarrow \overline{e}_{\alpha}+e_{\beta}$ via the triangular loop consisting of two $N_{i}$ and one $\phi^{-}$, and $\phi^{0}\rightarrow \phi^{+}+\phi^{-}$ via the triangular loop consisting of two $N_{i}$ and one $e_{\alpha}$. Note that $\phi^{0}$ can not decay into the final states of $N_{2,3}$ due to (8). However, the decay rate of (i) is larger than ones of (ii) and (iii) for the parameter values $Y_{N_{1}}^{2}\sim10^{-8}$, $Y_{N_{2,3}}^{2}\sim10^{-2}$, and $(Y_{D}Y_{D}^{\dagger})_{ij}\sim10^{-2}$, so the total decay width of $\phi^{0}$ is mainly dominated by (i).

 The process of Figure 2 explicitly violates ``$-2$" unit of the $B-L$ number. In addition, the decay rate of $\phi^{0}\rightarrow N_{1}+\overline{e}_{\alpha}+\phi^{-}$ is different from one of its $CP$-conjugate process $\phi^{0}\rightarrow \overline{N}_{1}+e_{\alpha}+\phi^{+}$ due to the interference between the tree diagram and the loop one. The $CP$ asymmetry of the two decay rates is defined and calculated as follows,
\begin{alignat}{1}
 &\varepsilon=\frac{\Gamma^{+}-\Gamma^{-}}{\Gamma_{\phi^{0}}}=\frac{1}{16\pi^{3}}\sum\limits_{i=2,3}Im[(Y_{D}Y_{D}^{\dagger})_{i1}^{2}]\frac{M_{N_{1}}M_{N_{i}}}{M^{2}_{\phi^{0}}}\left(ln\frac{M_{\phi^{0}}}{M_{\phi^{-}}}-\frac{5}{4}+\frac{M^{2}_{\phi^{0}}}{8M^{2}_{N_{i}}}+\frac{M^{4}_{\phi^{0}}}{24M^{4}_{N_{i}}}\right), \nonumber\\
 &\Gamma_{\phi^{0}}\approx\Gamma(\phi^{0}\rightarrow N_{1}+N_{1})+\Gamma(\phi^{0}\rightarrow \overline{N}_{1}+\overline{N}_{1})=\frac{M_{\phi^{0}}Y^{2}_{N_{1}}}{32\pi}\,, \nonumber\\
 &\Gamma^{\pm}=\sum\limits_{\alpha}\Gamma(\frac{\phi^{0}\rightarrow N_{1}+\overline{e}_{\alpha}+\phi^{-}}{\phi^{0}\rightarrow \overline{N}_{1}+e_{\alpha}+\phi^{+}})=\Gamma_{tree}+\Gamma_{loop}^{\pm}\,, \nonumber\\
 &\Gamma_{tree}=\frac{M_{\phi^{0}}Y^{2}_{N_{1}}(Y_{D}Y_{D}^{\dagger})_{11}}{(8\pi)^{3}}(ln\frac{M_{\phi^{0}}}{M_{\phi^{-}}}-\frac{3}{2}),
\end{alignat}
where the first and last two terms in $\varepsilon$ respectively arise from (b) and (c) in Figure 2. As a result of the Dalitz limit $M^{2}_{\phi^{-}}+M^{2}_{e_{\alpha}}<(M_{\phi^{-}}+M_{e_{\alpha}})^{2}\leqslant s_{12}=(p_{\overline{e}_{\alpha}}+p_{\phi^{-}})^{2}$, the imaginary part of the loop integration factor of (b) is derived from the three-point function $Im[C_{11}(M^{2}_{\phi^{-}},M^{2}_{e_{\alpha}},s_{12},M^{2}_{e_{\beta}},M^{2}_{N_{i}},M^{2}_{\phi^{-}})]=\frac{2\pi i}{s_{12}}$, while the imaginary part of (c) is derived from the two-point function $Im[B_{1}(s_{12},M^{2}_{e_{\beta}},M^{2}_{\phi^{-}})]=(1-\frac{M^{2}_{\phi^{-}}}{s_{12}})\pi i$. Obviously, the $CP$ asymmetry $\varepsilon$ is non-vanishing on account of the $CP$-violating phases in $Y_{D}$. We can learn from (8) the rough ratios $\frac{M_{\phi^{-}}}{M_{\phi^{0}}}\sim 10^{-4}$, $\frac{M_{\phi^{0}}}{M_{N_{2,3}}}\lesssim1$, and $\frac{M_{N_{1}}M_{N_{2,3}}}{M^{2}_{\phi^{0}}}\sim 10^{-4}$, then we can easily estimate $\varepsilon\sim 10^{-8}$ provided $Im[(Y_{D}Y_{D}^{\dagger})_{1i}^{2}]\sim10^{-3}$, which is a reasonable and typical value. In fact, the self-energy contribution to $\varepsilon$ is relatively small, we can safely ignore it.

  A simple calculation shows that the decay rate $\Gamma^{\pm}$ in (12) is smaller than the universe expansion rate, namely
\ba
 \Gamma^{\pm}\approx\Gamma_{tree}<H(M_{\phi^{0}})=\frac{1.66\sqrt{g_{*}}M^{2}_{\phi^{0}}}{M_{Pl}}\,,
\ea
therefore the decay process of Figure 2 is actually out-of-equilibrium. At the temperature of $M_{\phi^{0}}$ the relativistic states include $N_{1}$ and $\phi^{-}$ besides all of the SM particles, so $g_{*}=110.5$ in (13).

  We have demonstrated that the decay process of Figure 2 completely satisfies Sakharov's three conditions \cite{14}, as a result, it can surely generate a $B-L$ asymmetry. This is given by the relation \cite{15},
\ba
 Y_{B-L}=\frac{n_{B-L}-\overline{n}_{B-L}}{s}=\kappa\frac{(-2)\varepsilon}{g_{*}}\,,
\ea
where $s$ is the entropy density and $\kappa$ is a dilution factor. If the decay is severe departure from thermal equilibrium, the dilution effect is very weak, then we can take $\kappa\approx1$.

  The above $B-L$ asymmetry is generated at the temperature of $M_{\phi^{0}}\sim1000$ TeV, therefore the sphaleron process is fully put into effect \cite{16}, it can convert a part of the $B-L$ asymmetry into the baryon asymmetry. This is expressed by the relation
\ba
 \eta_{B}=\frac{n_{B}-\overline{n}_{B}}{n_{\gamma}}=7.04\,c_{s}Y_{B-L}\approx 6.15\times10^{-10},
\ea
where $c_{s}=\frac{8N_{f}(N_{f}+1)}{22N_{f}^{2}+81N_{f}+42}=\frac{32}{161}$ (for $N_{f}=3$) is the sphaleron conversion coefficient in the model. $7.04$ is a ratio of the entropy density to the photon number density. $6.15\times10^{-10}$ is the current value of the baryon asymmetry \cite{17}. Note that $\Delta$ and $\phi^{\mp}$ participate in the sphaleron process besides the SM particles, but $\phi^{0}$ and $N_{i}$ are not involved in it since they are singlets under the $G_{SM}$. When the universe temperature falls below the electroweak scale $\sim100$ GeV, the sphaleron process is closed and the baryon asymmetry is kept up to the present day.

\vspace{0.6cm}
\noindent\textbf{V. Numerical Results and Discussions}

\vspace{0.3cm}
  We concretely show some numerical results of the model. All of the SM parameters have been fixed by the current experimental data \cite{18}. The model only contains fewer parameters beyond the SM. For the sake of simplicity, we choose such typical values for some fundamental parameters as follows,
\begin{alignat}{1}
 &v_{\phi}=M_{\Delta}=2000\:\mathrm{TeV},\hspace{0.5cm} M_{\phi^{0}}=M_{N_{2}}=M_{N_{3}}=1000\:\mathrm{TeV}, \nonumber\\
 &v_{H}=246\:\mathrm{GeV},\hspace{0.5cm} M_{\phi^{-}}=300\:\mathrm{GeV},\hspace{0.5cm} M_{N_{1}}=248.2(267.2)\:\mathrm{GeV}, \nonumber\\
 &\lambda_{7}(Y_{\nu})_{33}=2\times10^{-9},\hspace{0.5cm} (Y_{D}Y_{D}^{\dagger})_{11}=0.01, \nonumber\\
 &Im[(Y_{D}Y_{D}^{\dagger})^{2}_{21}]=Im[(Y_{D}Y_{D}^{\dagger})^{2}_{31}]=0.0035(0.0033).
\end{alignat}
The above values are completely in accordance with the model requirements, in particular, satisfy the relation of (8). $\lambda_{7}(Y_{\nu})_{33}$ is determined by fitting the upper bound of neutrino mass (which is assumed as $m_{\nu_{3}}$). $M_{N_{1}}$ is determined by fitting the relic abundance of the CDM, but it has double-value solutions. $Im[(Y_{D}Y_{D}^{\dagger})^{2}_{i1}]$ is determined by fitting the baryon asymmetry, its double values correspond to the double values of $M_{N_{1}}$. The rest of the parameters are fixed. Now put (16) into the foregoing equations, we can correctly reproduce the desired results,
\ba
 m_{\nu_{3}}\approx 0.043\:\mathrm{eV},\hspace{0.5cm} \Omega_{N_{1}}h^{2}\approx 0.12\,,\hspace{0.5cm} \eta_{B}\approx 6.15\times10^{-10}.
\ea
These are exactly the current experimental data \cite{18}. In addition, we can work out $\frac{\Gamma_{tree}}{H}\approx0.09$ by use of (12) and (13), this demonstrates that the decay of Figure 2 is severe out-of- equilibrium indeed.

  Figure 3 shows the curves of $M_{N_{1}}$ versus $M_{\phi^{-}}$ for the three values of $(Y_{D}Y_{D}^{\dagger})_{11}$ while the other parameters are fixed by (16).
\begin{figure}
 \centering
 \includegraphics[totalheight=8cm]{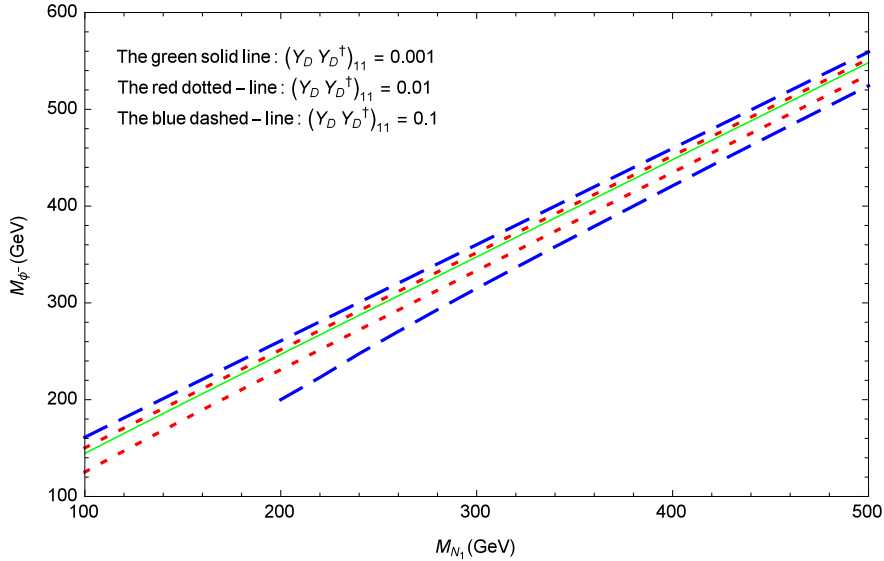}
 \caption{The curves of $M_{N_{1}}$ versus $M_{\phi^{-}}$ for the three values of $(Y_{D}Y_{D}^{\dagger})_{11}$ while the other parameters are fixed by (16), which can correctly fit $\Omega_{N_{1}}h^{2}\approx0.12$.}
\end{figure}
Any point of these curves can correctly fit $\Omega_{N_{1}}h^{2}\approx0.12$. For the case of $(Y_{D}Y_{D}^{\dagger})_{11}=0.1$, $M_{N_{1}}$ has only a single value when $M_{N_{1}}\lesssim200$ GeV. It can be seen from the curves that $\frac{M_{\phi^{-}}}{M_{N_{1}}}$ is almost invariant for those points in the same curve. The specific value of $\frac{M_{\phi^{-}}}{M_{N_{1}}}$ leads that (9) can give rise to a weak interaction cross-section, eventually, it brings about the correct $\Omega_{N_{1}}h^{2}\approx0.12$. However, the value areas of $M_{N_{1}}$ and $M_{\phi^{-}}$ are reasonable and moderate in Figure 3. The experimental search for $N_{1}$ and $\phi^{-}$ should focus on this parameter space.

  Finally, we simply discuss the test of the model at the colliders. The two particles of $N_{1}$ and $\phi^{-}$ are probably probed in the near future. The relevant processes are as follows,
\begin{alignat}{1}
 &e^{-}+e^{+}\rightarrow N_{1}+\overline{N}_{1},\hspace{0.5cm} e^{-}+e^{+}\:or\:p+\overline{p} \rightarrow\gamma\rightarrow \phi^{-}+\phi^{+}, \nonumber\\
 &p+p \rightarrow\gamma+\gamma\rightarrow \phi^{-}+\phi^{+},\hspace{0.5cm} \phi^{\mp}\rightarrow e_{\alpha}^{\mp}+N_{1}.
\end{alignat}
A pair of $N_{1}$ can be produced by $e^{-}+e^{+}$ via the t-channel mediation of $\phi^{-}$, but its cross-section is less than $\sim10^{-12}$ $\mathrm{GeV^{-2}}$. A pair of $\phi^{\mp}$ can be produced by $e^{-}+e^{+}$ or $p+\overline{p}$ via the s-channel gamma photon mediation if the center-of-mass energy is enough high. However, the future colliders as CEPC and ILC have some potentials to achieve these goals \cite{19}. At the present LHC \cite{20}, we also have a chance to find $\phi^{\mp}$ via two gamma photon fusion, furthermore, we can detect the CDM $N_{1}$ by the decay of $\phi^{\mp}$. Of course, this needs the researchers make a great effort. For test of leptogenesis mechanism of the model, the collider energy need be improved to 1000 TeV or so, then we possibly generate some asymmetric leptons in the laboratory. In short, it will be very large challenges to actualize these experiments, but the model tests possibly come true in the near future.

\vspace{0.6cm}
\noindent\textbf{VI. Conclusions}

\vspace{0.3cm}
  In summary, we only need to make the minimal extension of the SM with $U(1)_{B-L}\otimes Z_{2}$, then we can simultaneously account for the tiny neutrino mass, the CDM and the matter-antimatter asymmetry only by the fewer parameters. All of the new physics beyond the SM arise from the $U(1)_{B-L}$ violation at the energy scale about 1000 TeV. The model can naturally explain the ``WIMP Miracle" and elegantly achieve the leptogenesis. In addition, $N_{1}$ and $\phi^{-}$ in the model have such masses as a few hundred GeVs, they are very possible to be probed by the future collider experiments. However, these new physics of the model are very attractive and worth researching in depth.

\vspace{0.6cm}
 \noindent\textbf{Acknowledgements}

\vspace{0.3cm}
  I would like to thank my wife for her large helps. This research is supported by the Fundamental Research Funds for the Central Universities Grant No. WK2030040054.

\end{document}